\documentclass[aps,prl,twocolumn,superscriptaddress]{revtex4-1}
\usepackage{graphicx}
\usepackage{caption}
\DeclareCaptionLabelSeparator{dot}{. }
\makeatletter
\def\justified{
	\let\\\@normalcr
	\@rightskip\z@skip \rightskip\@rightskip
	\leftskip\z@skip
	\parindent 0em\relax
	\setlength{\parfillskip}{0pt plus 1fil}}
\DeclareCaptionJustification{justified}{\justified}
\usepackage{subcaption}
\usepackage{amsmath}
\usepackage{braket}
\usepackage{units}
\usepackage{ragged2e}
\usepackage[colorlinks,urlcolor=blue ,citecolor=blue ,linkcolor=blue ]{hyperref}
\usepackage{xcolor}
\usepackage{nicefrac} 
\usepackage{lipsum}
\usepackage{nameref}
\usepackage{hyperref}
\usepackage{textcomp} 
\usepackage{urwchancal}

\newcommand{\Er}{\ensuremath{^{166}}{\rm Er}}
\newcommand{\Dy}{\ensuremath{^{164}}{\rm Dy}}

\newcommand{\PC}{\mathcal{C}}

\newcommand{\as}{\ensuremath{a_{\rm s}}}

\newcommand{\tho}{t_{\rm h}}

\newcommand{\um}{\mu{\rm m}}

\newcommand{\krot}{k_{\rm rot}}

\newcommand{\rvec}{\mathbf r}

\newcommand{\deltaphi}{\delta\varphi}

\begin{document}
	
\title{Excitation spectrum of a trapped dipolar supersolid and its experimental evidence}
	
\author
{G.\,Natale}
\affiliation{
		Institut f\"ur Experimentalphysik, Universit\"at Innsbruck, Technikerstra{\ss}e 25, 6020 Innsbruck, Austria}
		
\author
{R.\,M.\,W.\,van\,Bijnen}
\affiliation{
		Institut f\"ur Quantenoptik und Quanteninformation, \"Osterreichische Akademie der Wissenschaften, Technikerstra{\ss}e 21a, 6020 Innsbruck, Austria}

\author
{A.\,Patscheider}
\affiliation{
		Institut f\"ur Experimentalphysik, Universit\"at Innsbruck, Technikerstra{\ss}e 25, 6020 Innsbruck, Austria}
		
\author
{D.\,Petter}
\affiliation{
		Institut f\"ur Experimentalphysik, Universit\"at Innsbruck, Technikerstra{\ss}e 25, 6020 Innsbruck, Austria}

\author
{M.\,J.\,Mark}
\affiliation{
		Institut f\"ur Experimentalphysik, Universit\"at Innsbruck, Technikerstra{\ss}e 25, 6020 Innsbruck, Austria}
\affiliation{
		Institut f\"ur Quantenoptik und Quanteninformation, \"Osterreichische Akademie der Wissenschaften, Technikerstra{\ss}e 21a, 6020 Innsbruck, Austria}	

\author
{L.\,Chomaz}
\affiliation{
		Institut f\"ur Experimentalphysik, Universit\"at Innsbruck, Technikerstra{\ss}e 25, 6020 Innsbruck, Austria}	

\author
{F.\,Ferlaino}
\affiliation{
		Institut f\"ur Experimentalphysik, Universit\"at Innsbruck, Technikerstra{\ss}e 25, 6020 Innsbruck, Austria}
\affiliation{
		Institut f\"ur Quantenoptik und Quanteninformation, \"Osterreichische Akademie der Wissenschaften, Technikerstra{\ss}e 21a, 6020 Innsbruck, Austria}	
	
\date{\today}
	
\begin{abstract}
We study the spectrum of elementary excitations of a dipolar Bose gas in a three-dimensional anisotropic trap across the superfluid-supersolid phase transition. Theoretically, we show that, when entering the supersolid phase, two distinct excitation branches appear, respectively associated with dominantly crystal and superfluid excitations. These results confirm infinite-system predictions, showing that finite-size effects play only a small qualitative role, and connect the two branches to the simultaneous occurrence of crystal and superfluid orders. Experimentally, we probe compressional excitations in an Er quantum gas across the phase diagram. While in the Bose-Einstein condensate regime the system exhibits an ordinary quadrupole oscillation, in the supersolid regime, we observe a striking two-frequency response of the system, related to the two spontaneously broken symmetries.

\end{abstract}

\maketitle
Supersolidity -- a paradoxical quantum phase of matter that combines crystal rigidity and superfluid flow -- was suggested more than half a century ago as a paradigmatic manifestation of a state in which two continuous symmetries are simultaneously broken~\cite{Boninsegni:2012}. In a supersolid, the spontaneously broken symmetries are the gauge symmetry, associated with the phase coherence in a superfluid, and the translational invariance, signalizing crystalline order. The striking aspect is that, in a supersolid of indistinguishable bosons, the same particles are participating in developing such two apparently antithetical, yet coexisting, orders. 
Originally predicted in quantum solids with mobile bosonic vacancies \cite{Andreev,Chester,Leggett}, the search for supersolidity has fueled research across different areas of quantum matter from condensed matter to atomic physics, including quantum gases with nonlocal interparticle interactions ~\cite{Henkel2010tdr, Cinti2010sdc, Boninsegni2012, Lu2015sds, Macia:2016, Cinti:2017, Wenzel:2017, Baillie:2018, Chomaz:2018, Ancilotto:2019, Boettcher:2019, Tanzi:2019, Chomaz:2019, Youssef2019,Zhang2019}. 

Recent experiments have revealed that axially elongated dipolar quantum gases can undergo a phase transition from a regular Bose-Einstein condensate (BEC), possessing a homogeneous density in the local-density-approximation sense, to a state with supersolid properties, where density modulation and global phase coherence coexist~\cite{Boettcher:2019,Tanzi:2019,Chomaz:2019}.  
Such experiments, complementing the ones with BECs coupled to light \cite{Leonard:2017, Li:2017, Leonard:2017b}, have opened a whole set of fundamental questions, covering the very real meaning of superfluidity in a supersolid state, its shear transport, and phase rigidity.

Of particular relevance is the study of the spectrum of elementary excitations, which governs the system’s response to perturbations~\cite{Landau47, Bogoliubov:1947,Pitaevskii:2016}. 
Typically, phase transitions occur in concomitance with drastic modifications of the excitation spectra,
-- as in the case of the emergence of roton excitations in He II or the phononic dispersion for BECs --
and similar dramatic changes are expected when crossing the superfluid-supersolid transition. Theoretical studies of uniform (infinite) gases with periodic boundary conditions and soft-core~\cite{saccani:2012, macri:2013, Rossotti2017} or dipolar interactions ~\cite{Macia:2012,Bombin:2017, Ancilotto:2019}, have shown two distinct branches appearing in the excitation spectrum of a supersolid state -- one for each broken symmetry. Their coexistence has been identified as an unambiguous proof of supersolidity, being the direct consequence of the simultaneous presence of superfluid and crystalline orders~\cite{Andreev, Pomeau:1994,saccani:2012, macri:2013}.

An important issue is to understand if these trademarks survive -- and can be measured -- in the experimentally relevant regimes of a finite-size quantum gas, confined in all three spatial dimensions. In this Letter, we address these points by performing full spectrum calculations and by experimentally exciting collective modes in an erbium quantum gas. Both theory and experiment show the existence of two distinct classes of excitations, one connected to crystal modes and the other to phase modes, providing the finite-size equivalent of the two-branches spectrum for infinite systems.

\begin{figure*}[t!]
	\includegraphics[width=\textwidth]{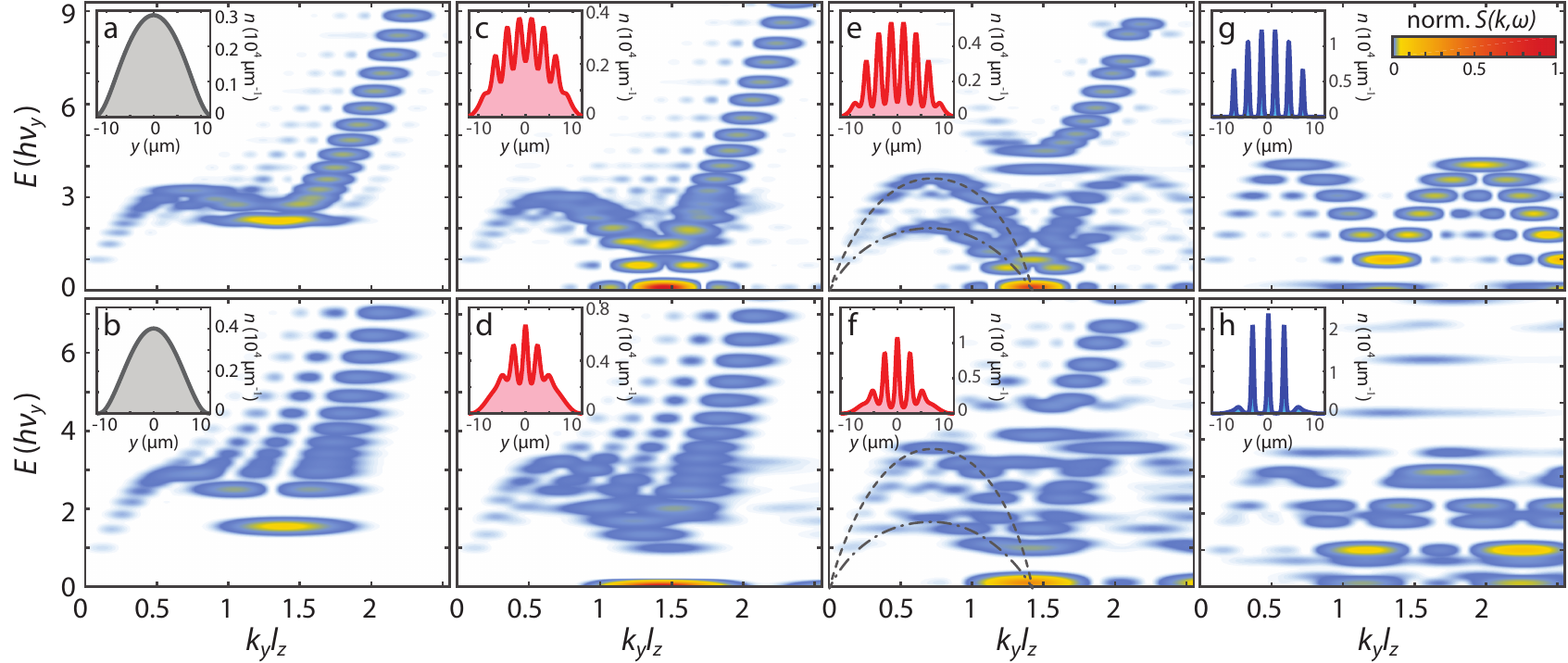}
	\caption {Axial excitation spectra of a trapped dipolar quantum gas across the BEC-supersolid-ID phase transition. The trap frequencies are $2\pi\times(260,\,29.6,\,171)$\,Hz. The upper (lower) row shows calculations for a \Dy~(\Er) quantum gas of $4\times 10^4$ ($5\times 10^4$) atoms in the BEC (a,b), supersolid (c-f) and ID (g,h) regimes, together with the corresponding ground-state density profiles (insets). (a,\,c,\,e,\,g) correspond to $\as=(92, 91, 90, 81)\,a_0$, and (b,\,d,\,f,\,h) to $\as=(50.8, 50.5, 50, 48)\,a_0$, respectively. In (e,\,f), the dashed and dash-dotted lines are guides to the eyes, indicating the two excitation branches. 
	The color map indicates the calculated DSF and $l_z$ is the harmonic oscillator length along the dipoles' direction.}
	
	 \label{fig:DSF} 
\end{figure*}

In our study, we consider a three-dimensional dipolar quantum gas confined in an axially elongated ($y$) harmonic trap with transverse orientation ($z$) of the atomic dipoles. 
These systems are well described by an extended Gross-Pitaevskii equation (eGPE), including nonlinear terms, accounting for contact interactions depending on the scattering length  $\as$, the anisotropic long-range dipole-dipole interaction (DDI), and 
quantum fluctuations in the form of a Lee-Huang-Yang type of correction ~\cite{Waechtler:2016,Bisset:2016,Waechtler:2016b,Schmitt:2016,Chomaz:2016,Baillie:2018,Ancilotto:2019,Tanzi:2019,Chomaz:2019,Boettcher:2019,Zhang2019}; see also \cite{Supmat}. We calculate ground-state wavefunctions, $\psi_0(\rvec)$, by minimizing the energy functional resulting from the eGPE using the conjugate-gradients technique~\cite{Ronen:2006}. As shown in Fig.\,\ref{fig:DSF} (insets), the ground state evolves with decreasing $\as$  from a regular BEC (a,b) to a supersolid state with axial density-wave modulation (c-f) and finally to an insulating array of independent droplets (ID) (g-h) \cite{Boninsegni2012, macri:2013, Ancilotto:2019, Boettcher:2019, Chomaz:2019}. 
 
The spectrum of elementary excitations is calculated by numerically solving the Bogoliubov de Gennes equations, which are obtained from an expansion of the macroscopic wavefunction as $\psi(\rvec, t)= \left[ \psi_0(\rvec) + \eta (u_l e^{-i\epsilon_l t/\hbar } + v_l^* e^{i\epsilon_l t/\hbar}) \right] e^{-i \mu t}$ with $\eta \ll 1$ and linearizing the eGPE around $\psi_0$~\cite{Pitaevskii:2016, Ronen:2006, Bijnen2010, Chomaz:2018}. Here, $\mu$ is the ground-state's chemical potential. By solving the resulting eigenvalue problem, we find a set of discrete modes, numbered by $l$, of energy $\epsilon_l = \hbar \omega_l$ and amplitudes $u_l$ and $v_l$. 
We calculate  the dynamic structure factor (DSF), $S(k,\omega)$,
which informs on the system's response when its density is perturbed at a given modulation momentum $k$ and with an energy $\hbar\omega$~\cite{Blakie:2002, Brunello:2001, Pitaevskii:2016}. 
Whereas in the absence of an external trap the spectrum is continuous and the DSF is a $\delta$-peak resonance at the Bogoliubov mode $(\omega_l, k_l)$, the confining potential yields instead a discretization of the excitation spectrum and a $k$ broadening in $S(k,\omega)$. For a given energy (i.\,e.\,a single mode), finite-size effects may even yield several peaks in $k$, see e.\,g.\,three-peak structures at large energy in Fig.\,1 (a,b). For the considered parameters, these finite-size effects are more pronounced in Er than Dy, since the latter  exhibits a larger number of maxima in the density-modulated phases, rendering its excitation spectrum more reminiscent of the infinite-system case; see Fig\,\ref{fig:DSF}.

Figure\,\ref{fig:DSF} shows the calculated excitation spectrum for ground-states in the regular BEC, the supersolid, and the ID phases for a Dy (upper row) and Er (lower row) quantum gas. 
In the BEC regime close to the supersolid transition (a,\,b), the spectrum of excitations shows a single excitation branch with the characteristic phonon-maxon-roton dispersion of a BEC~\cite{ODell2003,Santos:2003, Blakie:2012,Bisset:2013,Lasinio:2013}, as recently measured \cite{Petter:2019}.
When the roton fully softens (at $\as=\as^*$), the ground-state becomes density modulated with a wave number close to the roton one, $\krot$. Here, the excitation spectrum develops additional structures, marked by the appearance of nearly degenerate modes; (c,\,d). When lowering $\as$, we find that these modes start to separate in energy, where some harden and the others soften, and two excitation branches become visible; (e,\,f).
This result resembles that of infinite systems, where the broken translational and gauge symmetry are each associated with the appearance of one excitation branch~\cite{saccani:2012,macri:2013,Ancilotto:2019}. Additionally, we observe that the spectrum acquires a periodic structure, reminiscent of Brillouin zones in a crystal, with reciprocal lattice constant $k\simeq\krot$. Modes with an energy higher than the maxon (energy maximum at $k<\krot$) seem to have a single-droplet-excitation character, and they will be the subject of future investigations. 
When further decreasing $a_s<\as^*$, the lower-lying branch decreases both in energy and in DSF values, whereas the opposite occurs for the higher branch. Eventually, when reaching the ID regime, the lower branch progressively vanishes, underlying the disappearance of global superfluidity; (g,\,h).

\begin{figure}[t!]
	\includegraphics[width=1\columnwidth]{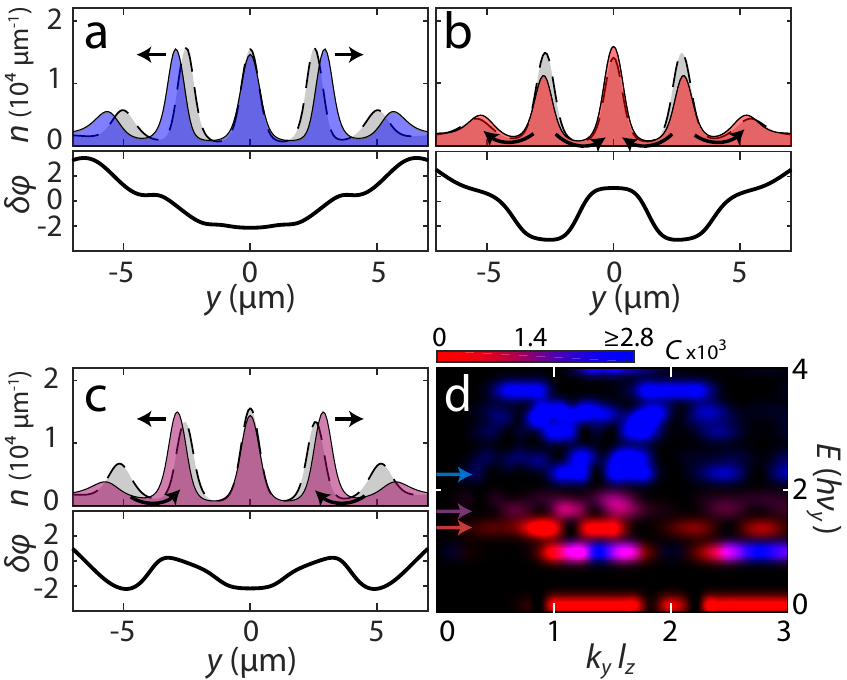}
	\caption{Evolution of three different even modes of the system calculated for $5\times 10^4$ Er atoms at $\as= 49.8\,a_0 $: (a) fourth (b) second and (c) third lowest lying even modes in energy with frequencies ($67.4\,$, $40.3\,$, $49.8\,$)\,Hz, corresponding to crystal, phase, and mixed modes, respectively. Each panel shows  $n=|\psi((0,y,0), t)|^2$ for $t=\pi/2\omega_l$  and $t = 3\pi/2\omega_l$ with $\eta = 0.15$ and the corresponding $\deltaphi(0,y,0)$. (d) DSF for the same setting as in (a-c) where the modes are colored according to their associated phase (red) or crystal (blue) character via $C$ \cite{Supmat}.
	}
	\label{fig:PhaseDensityMode} 
\end{figure}

We focus on the properties of the excitation spectrum in the supersolid regime. The interesting question is how the two branches relate to the two orders in the systems, crystal and superfluid. To gain insight, we study the system's dynamics when a single mode $l$ is excited with amplitude $\eta \ll 1$ by writing $\psi(t) e^{i\mu t} \approx \sqrt{|\psi_0|^2 + 2 \eta \delta \rho_l \cos \omega_l t}e^{-i \eta \deltaphi_l \sin \omega_l t}$, in terms of density perturbations $\delta \rho_l = (u_l + v_l^*) \left|\psi_0\right|$ and phase perturbations $\deltaphi_l = (u_l - v_l^*) / \left|\psi_0\right|$. 
The subsequent time evolution of the axial density profile is shown in Fig.\,\ref{fig:PhaseDensityMode}(a-c) for three relevant cases. For simplicity, only the two extremes of the mode oscillation are shown. 
The mode character can be understood by noting that phase gradients correspond to mass currents. Large gradients \textit{inside} a density peak imply motion of the density peak (e.g. Fig.\,\ref{fig:PhaseDensityMode}(a))
and relate to \emph{crystal modes}. Large phase gradients \textit{between} density peaks, signify a superfluid current of particles tunneling from one density peak to another (e.\,g.~Fig.\,\ref{fig:PhaseDensityMode}(b)), and are associated with \emph{phase modes}.
However, in our system, the phase/crystal mode classification is not strict and we find that these two characters mix; see Fig.\,\ref{fig:PhaseDensityMode}\,(a-c). Particularly, we observe both behaviors simultaneously in Fig.\,\ref{fig:PhaseDensityMode}\,(c). Such a mixing is expected from the long-range nature of the DDI, coupling density, and position of the peaks~\cite{saccani:2012, macri:2013}. Note that the character of the mode can change with $\as$. For instance, the mode in Fig.\ref{fig:PhaseDensityMode}\,(c) develops an almost pure crystal character for decreasing $\as$.
To quantify a mode's character, we plot in Fig.\,\ref{fig:PhaseDensityMode}\,(d) the DSF spectrum at a fixed $\as$, colored according to the ratio $C$ of phase variances inside, and between the density peaks \cite{Supmat}. This allows us to differentiate the dominant character of the two branches, being phase type for the lower branch, and crystal type for the upper one.

\begin{figure}[t!]
	\includegraphics[width=1\columnwidth]{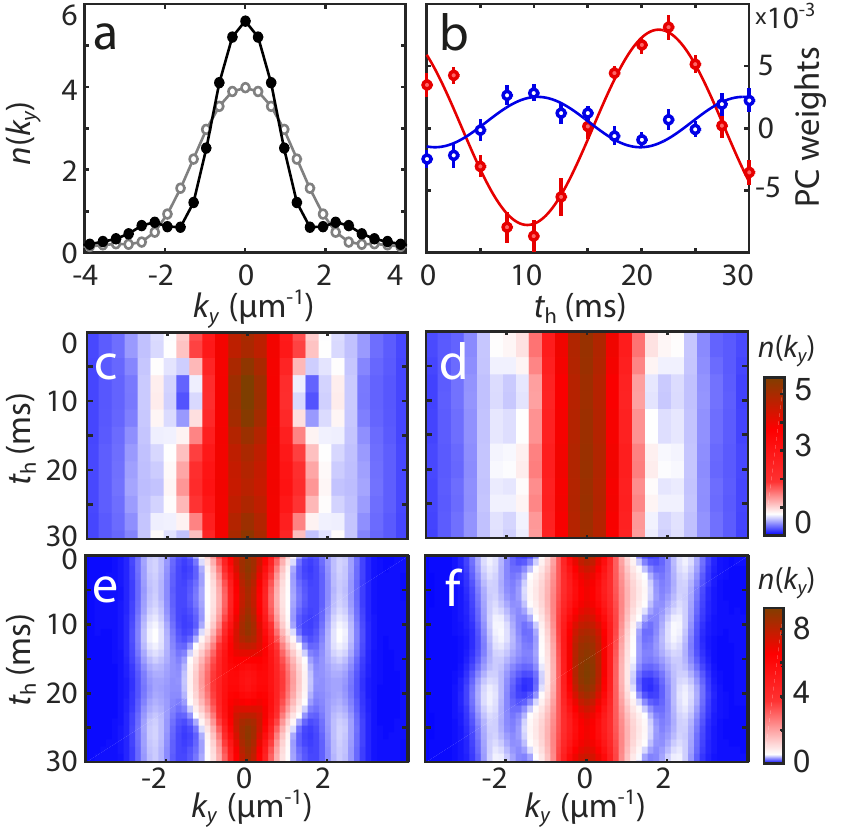}
	\caption{ (a) Example of a measured mean interference pattern in the renormalized central cut of the density distribution $n(k_y)$ for $\tho=5$\,ms in the supersolid regime at $\as=49.8\,a_0$ (filled circles) and in the BEC regime at $\as=51.7\,a_0$ (open circles). (b-d) PCA results at $\as=49.8\,a_0$. 
	(b) Time evolution of the weights of PC1 (filled circles) and PC2 (open circles) together with their sine fit. Error bars denote the standard error of the mean.  (c-d) Evolution of the partially recomposed $n(k_y)$ accounting for the population of PC1 (c) and PC2 (d) only.
	(e,\,f) Calculated time evolution of $n(k_y)$ from excitation of the mode shown in Fig.\,\ref{fig:PhaseDensityMode}\,(b) and (c), respectively, and using $\eta=0.15$.
	}
	\label{fig:PCAex} 
\end{figure}

To test our predictions, we experimentally study the collective excitations in an erbium quantum gas across the BEC-supersolid-ID phases. We prepare a BEC at $\as = 64\,a_0$. The atoms are confined in an axially elongated optical-dipole trap of harmonic frequencies $2\pi\times(\nu_x,\nu_y,\nu_z)=2\pi\times(259(2), 30 (1), 170(1))$\,Hz and polarized along $z$ by an external magnetic field; see \cite{Chomaz:2018,Chomaz:2019}. To probe our system, we perform standard absorption imaging after 30\,ms of time-of-flight expansion, yielding measurements of the momentum space density $n(k_x,k_y)$~\cite{Supmat}. Using the tunability of the contact interaction via magnetic Feshbach resonances \cite{Chin2010fri}, we can prepare the system at desired locations in the phase  diagram  – in the BEC, supersolid, or ID phase – by linearly ramping down $\as$ in 20\,ms to the target value. We then allow the system to stabilize for 10\,ms. At this point, we record an atom number of typically $5\times10^4$ for the supersolid regime.
We confirmed the relevant $\as$ ranges by repeating the matter-wave interferometric analysis of Ref.~\cite{Chomaz:2019}.
While in the BEC region the momentum distribution shows a regular, nearly Gaussian single peak, in the supersolid regime the in-trap density modulation gives rise to coherent interference patterns along $k_y$, consisting of a central peak with two lower-amplitude side peaks; see Fig.\,\ref{fig:PCAex}(a).

After preparing the system in the desired phase, we excite collective modes in the gas by suddenly reducing the axial harmonic confinement to 10$\%$ of its initial value (i.\,e.\,$\nu_y\approx3\,$Hz) for 1\,ms, before restoring it again. The atomic cloud is subsequently held for a variable time $\tho$, before releasing it from the trap and recording the time evolution of $n(k_x,k_y)$. As the lifetime of the supersolid state is limited to around 40\,ms~\cite{Chomaz:2019}, we focus on $\tho\leq30$\,ms. As expected, in the BEC phase, we predominantly observe an oscillation of the axial width, connected to the lowest-lying quadrupole mode~\cite{Pitaevskii:2016}.
In the supersolid regime, the situation is more complex; see Fig.\,\ref{fig:DSF}(c-f). Here, multiple modes, of both crystal and phase character, can be simultaneously populated, resulting in a convoluted dynamics of the interference pattern.

We therefore employ a {\em model-free} statistical approach, known as principal component analysis (PCA)~\cite{Jolliffe:2002}, to study the time evolution of the measured interference patterns at a fixed $\as$. This method has been successfully used to study e.\,g.,\,matter-wave interference~\cite{Segal:2010} and collective excitations~\cite{Dubessy:2014} in ultracold-gas experiments. The PCA analyzes the correlations between pixels in a set of images, decomposes them into statistically independent components, and orders these principal components (PCs) according to their contributions to the overall fluctuations in the dataset. 

In a dataset probing the system dynamics after an excitation, the PCA can identify the elementary modes with the PC’s weights in the individual images exhibiting oscillations at the mode  frequencies~\cite{Supmat, Dubessy:2014}. 
We apply the PCA to the time-evolution of the interference patterns after the trap excitation. Figure \ref{fig:PCAex}\,(b) shows the PCA results in the supersolid regime at $\as=49.8\,a_0$. We identify two leading PCs, which we label as PC1 and PC2. Their weights oscillate with different amplitudes and at distinct frequencies, namely 41(1)Hz for PC1 and 52(5)Hz for PC2. The comparison between the measured frequencies and the theoretically calculated mode energies indicates that, following our trap excitation, the second and third lowest lying even-mode are simultaneously populated. As shown in Fig.\,\ref{fig:PhaseDensityMode}\,(b) and (c), these modes possess a phase and a mixed character, respectively.  
Note that we apply an overall shift of $-4.3\,a_0$  to the $\as$  value for the experimental data; for more details see the discussion in Refs. \cite{Petter:2019,BoettcherQCorr:2019}. 

To visualize the role of each PC on the interference-pattern dynamics, we apply a partial recomposition of the images, accounting only for the PC of interest; see \cite{Supmat}.
The effect of PC1 on the axial dynamics is shown in Fig.\,\ref{fig:PCAex}(c), mainly being an axial breathing of the central peak, accompanied by weaker in-phase breathing of the side peaks. Instead, PC2 exhibits a dominant variation of the side-peak amplitude; see Fig.\,\ref{fig:PCAex}\,(d).
These results show a good agreement with the calculated time evolutions  of the interference patterns for the second and third even modes, shown in Fig.\,\ref{fig:PCAex}\,(e-f).

\begin{figure}[t!]
	\includegraphics[width=1\columnwidth]{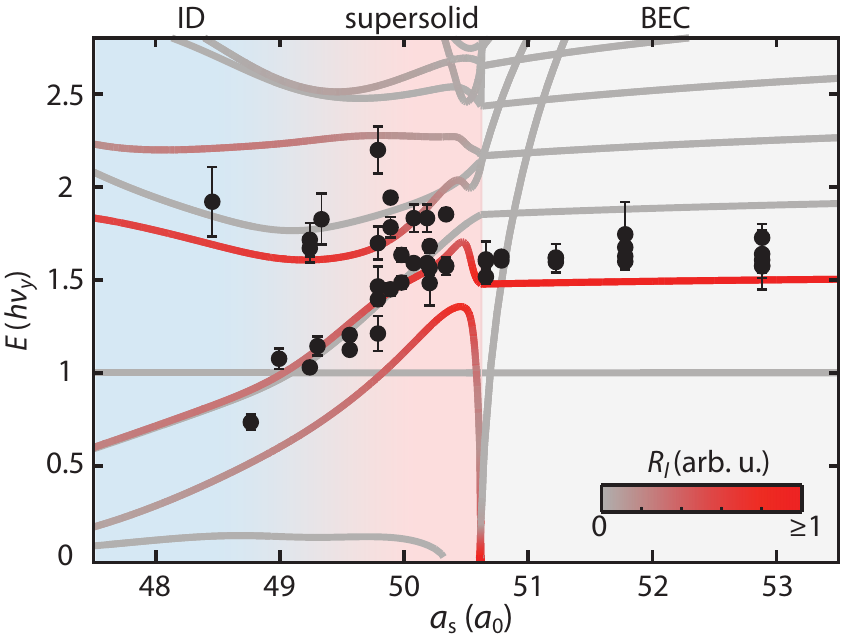}
	\caption{Comparison between the mode energy obtained from the theory calculations and the energies extracted from the PCs (circles). 	The gradual color code of the theory lines represents the relative strength of $R_l$ going from strong (red) to no (gray) coupling. Error bars denote one standard deviation from the fit. The background color indicates the BEC, supersolid, and ID regions  (see upper labels), identified using a matter-wave interferometric analysis of the experimental data \cite{Chomaz:2019}.}
	\label{fig:freqandampl_PC} 
\end{figure}

Finally, we study the evolution of the modes across the BEC to supersolid and ID phases.
We repeat the collective excitation measurements for various $\as$, and, using the PCA, we extract the oscillation frequencies of all the leading PCs.
Figure \ref{fig:freqandampl_PC} shows our experimental results together with the mode tracking from the spectrum calculations. For a give elementary mode $l$, we plot $\omega_l$ as well as the response amplitude $R_l=m\omega_y^2\langle \ell |\hat{y}^2|0 \rangle/2\hbar\omega_l$, which indicates the probability to be excited by our trap-excitation scheme. For completeness, the figure shows both even and odd modes, although only even modes are coupled to our trap-excitation scheme. Here, $|0 \rangle$ and $|\ell \rangle$ denote, respectively, the ground and excited states of interest, and $\hat{y}$ is the axial position operator.

In the BEC regime, besides the roton mode that progressively softens with decreasing $\as$, the other modes show a regular spacing in energy and are nearly constant with $\as$.  
In both the theory and experiment, we observe that just one mode couples to the trap-excitation scheme. This mode has a compressional, axial breathing character. Experimentally, we observe that all the leading PCs oscillate at the same frequency, suggesting that they account for the same mode~\cite{Supmat}. In this regime, both the PCs frequencies, $\omega_l$, and $R_l$ remain rather constant.
At the supersolid phase transition, reached around $\as = 50.6\,a_0$, the numerical calculations reveal that different modes undergo an abrupt change and can mix with each other. Their energy and phase/crystal character exhibits a strong dependence on $\as$. Here, several modes respond to the trap-excitation scheme, as shown by the value of $R_l$. 
In the PCA we observe that the leading PCs now oscillate at distinct frequencies and have different characters (see also Fig.\,\ref{fig:PCAex}). One set of PCs reduces their frequency when lowering $\as$, indicating (at least) one phase mode that softens strongly in the supersolid regime, even below the trap frequency $\nu_y$. Another set of PCs shows a frequency that remains hard when decreasing $\as$. Calculations of $C$ show that this mode changes character along the phase diagram and eventually becomes crystal type. 

In conclusion, the overall agreement between the experiment and theory confirms the calculations in the supersolid regime, revealing two distinct branches with respective crystal and superfluid characters. The trademarks of supersolidity expected in infinite systems thus carry over to the finite-size ones currently available in laboratories. The knowledge of the excitation spectrum will provide the base for future investigations related to the superfluid properties and phase rigidity in a supersolid state.

\begin{acknowledgments}
 
We thank D.\,Baillie, R.\,Bisset, B.\,Blakie, T.\,Pfau, A.\,Recati, L.\,Santos, and P.\,Silvi for stimulating discussions. We acknowledge useful conversations with the participants of the meeting on ``Perspectives for supersolidity in dipolar droplet arrays'' in Stuttgart, where we became aware of related work by the groups of G.\,Modugno~\cite{Tanzi2:2019} and T.\,Pfau~\cite{Guo:2019}. Part of the computational results presented have been achieved using the HPC infrastructure LEO of the University of Innsbruck. This work is financially supported through an ERC Consolidator Grant (RARE, no.\,681432) and a DFG/FWF (FOR 2247/PI2790), and the project PASQUANS of the EU Quantum Technology flagship.
\end{acknowledgments}

* Correspondence and requests for materials
should be addressed to Francesca.Ferlaino@uibk.ac.at.

%
\appendix
\renewcommand\thefigure{\thesection S\arabic{figure}}   
\setcounter{figure}{0}   

\subsection{Calculation of the Bogoliubov Spectrum}
Our theory is based on an extended version of the Gross-Pitaevskii equation (eGPE)
\begin{align} \label{eq:GPE}
{\mathrm i}\hbar\frac{\partial\psi(\rvec, t)}{\partial t}&=\Big(-\frac{\hbar^2 \nabla^2}{2m}+V(\rvec) + \int d\rvec' U({\mathbf r}-{\mathbf r}')n({\mathbf r}') \nonumber\\ & \hspace{2cm}+ \Delta \mu[n]\Big)\psi(\rvec, t),
\end{align}
where $\psi(\rvec, t)$ is the dipolar quantum-gas' wave function $\psi(\rvec, t)$. The eGPE includes the kinetic energy, external trap potential and the mean-field effect of the interactions \cite{Pitaevskii:2016,Baranov:2008}. 
The first three terms of Eq.\,\eqref{eq:GPE} account for the kinetic energy, the external harmonic trapping potential, and the mean-field interactions, respectively. The latter includes  the contact and the dipolar interactions. In order to study the supersolid phase, it is fundamental to also include  a beyond-mean-field corrections in order to stabilize the supersolid state against the roton instability. This is done by adding a term in the form of the Lee-Huang-Yang correction, $\Delta \mu[n]$~\cite{Waechtler:2016,Bisset:2016,Waechtler:2016b,Schmitt:2016,Chomaz:2016,Baillie:2018,Ancilotto:2019,Tanzi:2019,Chomaz:2019,Boettcher:2019,Zhang2019}; see also~\cite{Gammal:2000, Bulgac:2002, Lu2015sds, Petrov:2015}. This is typically included as a correction to the chemical potential obtained under the assumption of local density approximation~\cite{Pelster:2011, Lee1957mbp}. However, recent experimental results have raised the questions about the range of validity of such a treatment since quantitative disagreements at a level of few $\%$ have been observed when comparing the theory results with the experimental findings~\cite{Schmitt:2016, Chomaz:2018, Igor:2018, Cabrera2018, Petter:2019, BoettcherQCorr:2019}. To the best of our knowledge, this is still an open question, which will need future additional theoretical investigations. To compensate for this effect, throughout this letter, we shift $\as$ by $-4.3 a_0$.
To calculate the ground-state (GS) wave-function, $\psi_0(\rvec)$, we then minimize the energy functional resulting from the eGPE using the conjugate-gradients technique~\cite{Ronen:2006}.

In a next step, we study the Bogoliubov de Gennes (BdG) excitation spectrum of a dipolar Bose-Einstein condensate trapped in a harmonic cigar shaped potential~\cite{Pitaevskii:2016, Ronen:2006}. Our calculations are obtained by expanding the wavefunction $\psi(\rvec, t)$ around $\psi_0 (\rvec)$. Here, we write:
\begin{equation*}
\label{eq:BdG_fieldexp}
\psi(\rvec, t)= \left( \psi_0(\rvec) + \eta \delta\psi(\rvec, t) \right) e^{-i \mu t},
\end{equation*}
where $\eta \ll 1$, $\mu$ is the chemical potential of the ground state and
\begin{equation*}
\delta\psi(\rvec, t)= u_l e^{-i\epsilon_l t/\hbar } + v_l^* e^{i\epsilon_l t/\hbar}.
\end{equation*}
The spatial modes $u_l$ and $v_l$ are oscillating in time with the corresponding frequency $\omega_l = \epsilon_l / \hbar$. We then linearize the eGPE around $\psi_0$ at first order in $\eta$. By solving the set of coupled linear equations, we obtained the discrete modes, numbered by $l$, of energy $\epsilon_l$ and amplitudes $u_l$ and $v_l$. We define the (odd) even parity of the mode from their amplitude $u_l$ and $v_l$ being (anti-)symmetric in $y$. 

In order to illustrate the spectrum, we compute the dynamic structure factor (DSF), since it directly gives information about the density response of the system when perturbed at specific energies and momenta. At $T=0$ the DSF is defined as~\cite{Blakie:2002,Chomaz:2018}:

\begin{eqnarray}
\label{EqStrucFac}
S( \mathbf {k}, \omega') &=&  \sum_l \left| \int \mathrm{d}\rvec \left[ u_l^*(\rvec) +v_l^*(\rvec)\right]\mathrm{e}^{\mathrm{i} \mathbf {k}\cdot \rvec}\psi_0(\rvec) \right|^2 \times \nonumber \\ 
&\times &\delta(\omega' - \omega),
\end{eqnarray}

where the sum is over the different spatial modes and $\mathbf {k}$ is the wave vector. In Fig.\,1 and Fig.\,2  we plot the DSF of Eq.\,\eqref{EqStrucFac}. For better visualization, we use an energy broadening of $0.09\,h\nu_y$ and $0.12\,h\nu_y$ for Fig.\,1 and  Fig.\,2, respectively, similar to what was done in Ref.~\cite{Blakie:2002}.

\subsection{Defining the mode character}
Within the Bogoliubov theory and in the linear regime, the effect of the population of the mode $l$ on the global state dynamics can be studied using the following expression \cite{Pitaevskii:2016}
\begin{equation*} \label{eq:psit} 
\psi(\rvec,t) e^{i\mu t} \approx \sqrt{|\psi_0(\rvec)|^2 + 2 \eta \delta \rho_l(\rvec) \cos \omega_l t}e^{-i \eta \deltaphi_l(\rvec) \sin \omega_l t}, 
\end{equation*} 
where the density fluctuations $\delta \rho_l = (u_l + v_l^*) \left|\psi_0\right|$ and phase fluctuations $\deltaphi_l = (u_l - v_l^*) / \left|\psi_0\right|$ have been separated.

 In order to evaluate the dominant character of each mode $l$, we introduce the quantity $C$.  As discussed in the main text, the crystal and phase mode differentiate from each other by the spatial region where $\deltaphi_l$ varies the most. For crystal modes, this is inside the density peaks, resulting e.\,g.\, in a center-of-mass motion of one individual peak, which leads to a change of the crystal structure. Differently, for phase modes, $\deltaphi_l$ changes the most between neighboring peaks, signalizing a particle exchange  between peaks and thus a modification of the atom numbers in the peaks.
 We quantify these two types of character by computing the spatial variance of $\deltaphi_l (\rvec)$ inside the density peaks, $V_{\textrm{in}}$, and in between them, $V_{\textrm{out}}$. 
 The quantities $V_{\textrm{in}}$ and $V_{\textrm{out}}$ are defined as follow.
 
For a given axial density cut of the GS wave function $|\psi_0(0,y,0)|^2$, we first define the region inside (between) the density peaks  by identifying the different density maxima (minima) and number them by $j\in[1, N_{\textrm{in(out)}}]$. In a next step, we compute the mean distance $d$ between all density minima to their neighbouring maxima. Finally, we isolate the region $R_j =[-d/3,+d/3]$ of space centered around each maxima (minima) and calculate:

\begin{eqnarray}
 V_{\textrm{in(out)}} &=&  \frac{1}{N_{\textrm{in(out)}}} \times \nonumber \\
  &\times &  \sum_{j=1}^{N_{\textrm{in(out)}}}  \langle |\deltaphi(0,y,0) - \langle \deltaphi(0,y,0)\rangle_{R_j}|^2 \rangle_{R_j}. \nonumber
\end{eqnarray}

\begin{figure}[t!]
	\includegraphics[width=1\columnwidth]{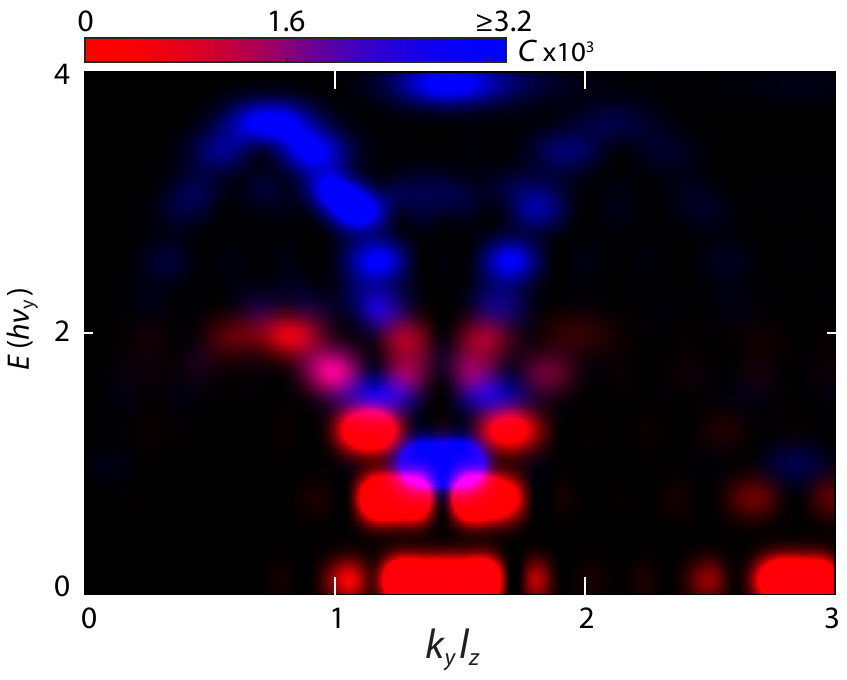}
	\caption{Characterization of the excitation modes for $N=4\times10^4$ atoms of \Dy \,at $a_s=90\,a_0$ in a trap of frequencies $2\pi\times(260,\,29.6,\,171)$\,Hz. As in the Er case (Fig.\,2\,(d) of the main text), the blue color reveals the dominant crystal character for the upper branch, whereas the red color shows the dominant phase character for the lower branch.}
	\label{figSuppMat:ModeCharactDy} 
\end{figure}

The mean $\langle \cdot\cdot\cdot \rangle_{R_j}$ is defined for a generic function $f$ as
\begin{align*}
\langle f(y) \rangle_{R_j} =\int_{y\in R_j} f(y)\,dy \bigg/ \int_{y\in R_j}\,dy.
\end{align*}
The mode character is then evaluated by considering the  ratio $C=\nicefrac{V_{\textrm{in}}}{V_\textrm{{out}}}$. $C$ is large for modes with prevalent crystal and small for the ones with dominant phase character.
In Fig.\,2\,(d) we encode the information on $C$ as a color scale on the DSF spectrum. The same color map is used to illustrate the modes of the panels (a-c) in Fig.\,2, confirming their correct assignment. For completeness, we also illustrate in Fig.\,\ref{figSuppMat:ModeCharactDy} the modes' character on the spectrum of a \Dy supersolid, using the parameters of Fig.\,1\,(e) of the main text.

\subsection{Applying the principal component analysis to our data}

\subsubsection{Dataset for applying the PCA}
To identify the excited modes from our experimental data, we apply a general statistical method called principal component analysis (PCA)~\cite{Jolliffe:2002,Segal:2010,Dubessy:2014} to a set of measured density distributions after a time-of-flight expansion.
For our trap-excitation measurement, a dataset for the PCA is composed as follow. For each target value of $\as$, we record 
the time evolution of the density distribution for holding time, $\tho$, between 0 and 30\,ms. For each $\tho$, we record between 15 and 30 repeated images, all together yielding a dataset of $N_m \gtrsim 200$ images. 
Each experimental run $i$ yields a two-dimensional density distribution $n_i(k_x,k_y)$. By performing a simple two dimensional Gaussian fit, we extract  $71\times71$ pixels region-of-interest (ROI) centered on  the atomic cloud (the pixel's width in $k_{x,y}$ is $0.32\um^{-1}$).  In addition, we post-select the shots in which the atom number, the axial cloud size and the transverse cloud size vary by less than 20\%, 30\% and 15\% than their mean values, respectively. 

\subsubsection{PCA's working principle}
\begin{figure}[t!]
	\includegraphics[width=1\columnwidth]{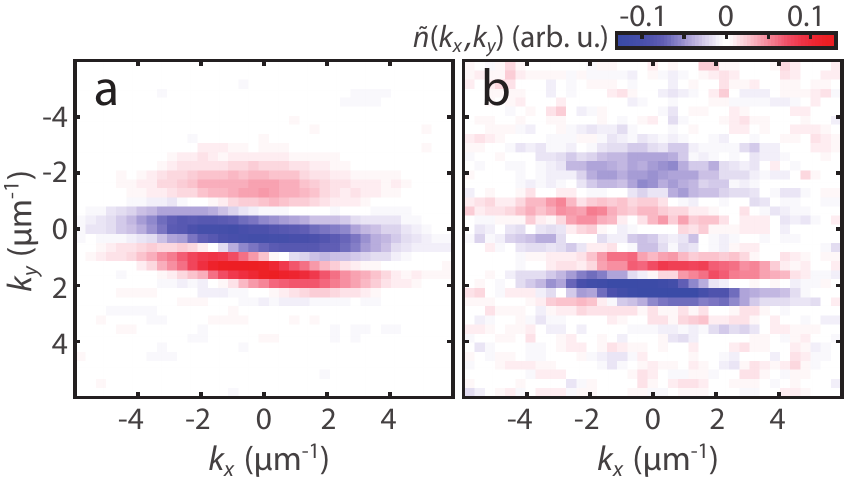}
	\caption{Examples of the two leading PCs for our dataset at $\as=50\,a_0$. (a) PC1 reveals a dominant fluctuation of the interference patterns in the central peak at $k_y\approx0\,\mu$m$^{-1}$  (central blue region) with a slighter change of the sidepeaks at $k_y\approx\pm2\,\mu$m$^{-1}$ (red regions) . (b) PC2 shows fluctuations in the interference patterns' sidepeaks around $k_y\approx2\,\mu$m$^{-1}$ and no significant change of the central peak.}
	\label{figSuppMat:ExamplePCs} 
\end{figure}

To apply the PCA, we represent each ROI of a dataset as a vector $\rho_i(s)$ where $s$ represent the index of the pixel ($s\in [1,N_p]$, $N_p$ is the number of pixels in one image). We compute the mean vector image $\bar\rho (s)= \sum_{i=1}^{N_m}\rho_i(s)/N_m$ and consider the variations of the pixel values in each vector image compared to $\bar\rho$, $\delta\rho_i(s)=\rho_i(s)-\bar\rho(s)$. Finally, we consider the covariance matrix of these variations $Cov(p,s)=\sum_{i=1}^{N_m}\rho_i(s)\rho_i(p)/(N_m-1)$, which is real symmetric. 
By simply diagonalizing the covariance matrix, the PCA constructs a new basis of $N_p$ vector-images, called principal components (PCs) and written $\PC_p(s)$ ($p\in [1,N_p]$) in the original pixel basis, that are uncorrelated one from an other. The PCs satisfy $Cov\PC_p= \lambda_p\PC_p$ where $\lambda_p$ is the eigenvalue of the covariance matrix associated to the PC $p$. The original vector images can be all rewritten in this new basis as $\rho_i(s)=\bar\rho(s)+ \sum_{p=1}^{N_s}w_{p,i}\PC_p(s)$,
where $w_{p,i}=\sum_{s=1}^{N_p}\PC_p(s)\rho_i(s)$ is the weight of the component $p$. We note that, by converting back the pixel representation to the original two-dimensional momentum space, the above decomposition means
\begin{equation}
\label{eq:PCA}
n_i(k_x,k_y)=\bar n(k_x,k_y)+ \sum_{p=1}^{N_s}w_{p,i}\PC_p(k_x,k_y),
\end{equation}
where $\PC_p(k_x,k_y)$ encompasses now the density-distribution change induced by the PC $p$. The fact that the covariance matrix is diagonal in the PC basis indicates that the PCs correspond to uncorrelated sources of variations in the dataset. More explicitly, the coefficients $w_{p,i}$ show no correlations in between different $p$. This feature makes the PCA a powerful tool, e.\,g.\,to identify and discriminate between elementary modes of different frequencies when applied to time evolution data, as used in Ref.\,\cite{Dubessy:2014}. An example of the obtained two leading PCs in the supersolid region is given in Fig.\,\ref{figSuppMat:ExamplePCs}.

\subsubsection{Identifying the elementary modes of a quantum gas via the PCA}

We quickly remind the working principle, of the identification of modes via the PCA. In the linear regime, the contribution of each mode to density oscillations is expected to decouple and separate temporal and spatial variations as:
\begin{equation}
\label{eq:psitall}
  n(\rvec,t) \approx n_0(\rvec,t)+ 2\sum_l \eta \delta \rho_l(\rvec) \cos\left(\omega_l t+\phi_l\right),  
\end{equation}
with $\phi_l$ an arbitrary phase for the mode $l$. This relation should also hold for the density distribution after the gas's free-expansion. If one considers that the image index $i$ encloses a time dependence ($t_i$), the equations \eqref{eq:PCA} and \eqref{eq:psitall} have a very similar structure, associating $\PC_p(k_x,k_y)$ and $w_{p,i}$ to $\rho_l(\rvec)$ and $\cos\left(\omega_l t_i\right)$, respectively. 
Thus the PCA-based identification of uncorrelated components in the time-evolution of the density profiles should enable to identify the elementary modes of the system. The corresponding PCs' weights are then expected to oscillate in time at the frequency $\omega_l$ of the modes. In particular, the PCA should separate the modes oscillating at different frequencies and differentiate them from other sources of fluctuations or of dynamics (e.g. dissipation). Following Ref.\,\cite{Dubessy:2014}, we note that modes can be properly distinguished if the period associated to their beating is smaller than the total time for which the time-evolution is recorded, or, even for shorter probe time, if they have different enough amplitudes of oscillations (i.e. excitation probability). 

From our dataset with repeated realizations of each hold time $\tho$, we thus consider, for each PC $p$, the mean weights at time $\tho$, $W_p(\tho)=\sum_{i/ t_i=\tho} w_{p,i}/ \sum_{i/ t_i=\tho}1$. We then fit $W_p(\tho)$ to a sine function $A_0+A_s\cos\left(\omega \tho+\phi\right)$ and extract the PC's frequency ($\omega$) and amplitude $A_s$ of oscillation. We then consider as relevant the PCs that show oscillation of amplitude $A_s>8\times 10^{-4}$, frequency $\nu >20\,$Hz, and where the oscillation frequency can be extracted with a precision $<10\%$.
Examples of the time evolution of $W_p$ and of their fits are shown in Fig.\,3\,(b).

We note that the PCA does not always assemble in a unique PC all the correlations in the pixel values that follow the same time dependence, and a single mode can be artificially split into several components in the analysis process. 
To better understand this behavior, we performed tests on theoretical calculations and compare them to the experiments. Theoretically, we specifically populate a single Bogoliubov mode on top of the ground-state, we then compute the interference patterns as a function of the hold time $\tho$, similar to what is done in Fig.\,3\,(e-f), and finally we apply the PCA. For each mode considered, both for regular (BEC) or density-modulated ground-states, several leading PCs are found to oscillate. Their frequencies match the mode frequency while their oscillation amplitude decreases with the PC's index. Typically the ratio in the oscillation amplitudes between the first and the second PC is about 10, and the amplitude of the larger-index PCs are negligible. Therefore, in the cases where the modes are the most strongly excited, i.e. mainly in the BEC regime (see $R_l$ scaling in Fig.\,4 for the excitation amplitude), one can indeed expect that several PCs are sensitive to a single mode in experiment, matching our observation.

From those theory tests, we can also better understand the origin of this artificial splitting of one mode in several PCs. Indeed, the oscillations of the different PCs are  found to have the same frequencies but different phases, typically shifted by about $\pi/2$. As it treats the pixels independently, the PCA gets confused by such $\pi/2$ phase shifts in the oscillations occurring in different regions of space, i.e. pixels' values that distinctly oscillate, starting from their extremal or medial values. The PCA then artificially splits the oscillations occurring in these different regions into several components while they correspond to the same mode.
Finally, this effect can be further favored in typically imperfect experimental settings, by the addition of experimental noise as well as other technical (e.g. imaging artifacts) or physical (e.g. dissipation) effects, which yield differences in the pixel values. Our theory tests, however, show that those additions are not the main reasons for the observed splitting.

Based on the conclusions of those tests, in the experiment (see discussion of Fig.\,4 of the main text), we interpret as probing distinct modes only the PCs showing different frequencies, while PCs whose frequencies match within their error bars are interpreted as probing a single elementary excitation of the system.

\subsubsection{Partial recomposition}

To isolate the effect of each PC on the complex time-evolution of the interference patterns, we use partial recomposition of the images inspired from Eq.\,\eqref{eq:PCA}. In particular we define
\begin{equation}
\label{eq:PCAp}
n^{(p)}(k_x,k_y,t)=\bar n(k_x,k_y)+ W_{p}(t)\PC_p(k_x,k_y).
\end{equation}
This is equivalent to consider that a single PC is "excited", similarly to what can be done in theory for the individual excited modes of the BdG spectrum (see Fig.\,2) and its description in the main text and Supp. Mat.). In Fig.\,3\,(c-d), we show examples of the axial cuts of $n^{(p)}(k_x,k_y,t)$ for two of the leading PCs. 
We note that here, as well as for all experimental data shown in this manuscript, the axial cuts correspond to the average of the density distributions for $|k_x|<1.6\,\um^{-1}$.

\end{document}